 \documentclass[final,3p,times,twocolumn]{elsarticle}

\usepackage{amsmath,amssymb}

\usepackage{subfigure}          

 \makeatletter
\def\ps@pprintTitle{%
  \let\@oddhead\@empty
  \let\@evenhead\@empty
  \def\@oddfoot{\reset@font\hfil\thepage\hfil}
  \let\@evenfoot\@oddfoot
}

\journal{Nuclear Physics A}

\begin{document}


\begin{frontmatter}



\title{Proton-proton cross-sections: the interplay between density and radius}

 \author[LIP]{R. Concei\c{c}\~{a}o\corref{cor1}}
 \ead{ruben@lip.pt}
 \cortext[cor1]{Corresponding author}
 \author[CENTRA,IST]{J. Dias de Deus}
 \author[LIP,IST]{M. Pimenta}
 \address[LIP]{LIP, Av. Elias Garcia, 14-1, 1000-149 Lisboa, Portugal}
 \address[IST]{Departamento de F\'{i}sica, IST, Av. Rovisco Pais, 1049-001 Lisboa, Portugal}
 \address[CENTRA]{CENTRA, IST, Av. Rovisco Pais, 1049-001 Lisboa, Portugal}

\begin{abstract}
We argue that there are two mechanisms responsible for the growth of high energy cross-sections $-$ $\sigma_{tot}$ and  $\sigma_{el}$, say $-$ in pp collisions. One is by the increase of matter \emph{density}, resulting in the black disk saturation. The other is exclusively by radial expansion affecting the periphery of the overlap area. Within this simple model we can reproduce all available data in pp from ISR to LHC.
In order to achieve a fast growth in the very high energy cosmic ray energies, we propose a fast black disk saturation followed by the dominance of geometric scaling (GS).
\end{abstract}

\begin{keyword}
Proton-proton cross-sections \sep Effective Radius \sep Density \sep Grey disk model

\end{keyword}

\end{frontmatter}



At high energy, or at high density, saturation occurs as a consequence of phase-space overcrowding of QCD matter \cite{ref1,ref2,ref3_1,ref3_2,ref3_3,ref3_4}. Unitary bounds may as well impose saturation and the possible reduction of particle multiplicities \cite{ref4}. As an example we have the black disk limit \cite{ref5},

\begin{equation}
\text{Im} F(b,s) \leq 1,
\label{eq:one}
\end{equation} 

where $Im F(b,s)$ is the imaginary part of the elastic amplitude $F(b,s)$, $b$ being the impact parameter and $s$ the square of the centre of mass energy.

\begin{figure}[htbp]
  \begin{center}
   \includegraphics[width=0.5\textwidth]{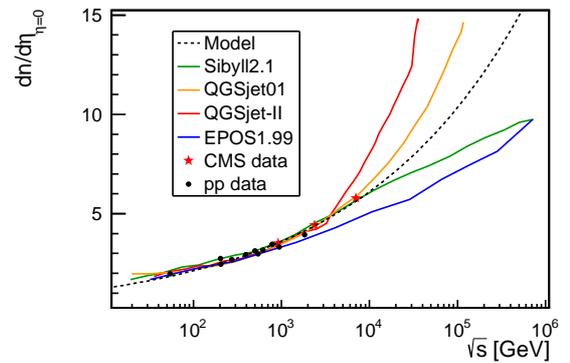}
   \caption{Charged multiplicity at pseudorapidity $\eta = 0$ \cite{UA1,UA5,CDF,CMS}. The colored lines show the evolution for different high energy hadronic interaction models \cite{QGS01,QGSII,Sibyll,EPOS} while the dotted line is obtained from the fit to the equation (\ref{eq:three}).}
   \label{fig:1}
 \end{center}
\end{figure}

In the string percolation model - see \cite{ref5} for a comparison between string percolation and the color glass condensate -  we have for the average number of elementary collisions,

\begin{equation}
\overline{\nu} = e^{2 \lambda Y},
\label{eq:two}
\end{equation}

where $Y$ is the beam rapidity, $Y=\ln \left( \sqrt{s} / m_p \right)$, and $\lambda$ a parameter, $0.2 \leq \lambda \leq 0.3$, related to deep inelastic scattering \cite{ref6} and to energy-momentum conservation \cite{ref7}. Note that in (\ref{eq:two}) we assume that as $Y \rightarrow 0$, $\overline{\nu} \rightarrow 1$. However due to string percolation \cite{percolation} the particle density is not proportional to $\overline{\nu}$ but rather proportional to $\sqrt{\nu}$,

\begin{equation}
\left. \frac{dn}{d\eta} \right| _{\eta=0} = a e^{\lambda Y},
\label{eq:three}
\end{equation}

where $a$ depends on the nature of produced particles, with $a \approx 1$ for charged particle production. In figure \ref{fig:1} we test the validity of (\ref{eq:three}) comparing it with $pp(\overline{\text{p}})$ data, from $\sqrt{s} \sim 10$ GeV to LHC energy ($\sqrt{s} = 7$ TeV) \cite{UA1,UA5,CDF,CMS} as well as with the most used high energy hadronic interaction models \cite{QGS01,QGSII,Sibyll,EPOS,dndeta0data1}. We obtain $a = 0.8$ and $\lambda = 0.23$.

We want next to relate the number of collisions and particle densities to $\text{Im} F(b,s)$ and cross-sections making use of a grey disk model and neglecting real part contributions:


\begin{equation}
\label{eq:greydisktot}
\sigma_{tot} (s)=  2 \pi \int d^2 b \text{Im} F(s,b) \rightarrow 2 \pi (1 - e^{- \overline{\Omega}(s)}) R^2(s), 
\end{equation}

\begin{equation}
\label{eq:greydiskcel}
\sigma_{el} (s)=  \pi \int d^2 b \left( \text{Im} F(s,b) \right)^2 \rightarrow \pi (1 - e^{- \overline{\Omega}(s)})^2 R^2(s), 
\end{equation}

\begin{equation}
\label{eq:greydiskinel}
\sigma_{inel}(s)=  \sigma_{tot}(s) - \sigma_{el}(s) = \pi (1 - e^{- 2 \overline{\Omega}(s)}) R^2(s),
\end{equation}

where $\overline{\Omega}(s)$ is the \emph{opacity} and $R(s)$ is the effective cross-section \emph{radius}. Note that $0 \leq  \overline{\Omega} \leq \infty$, or, $0 \leq  F(s,b) \leq 1$, and that $\overline{\Omega}$ is averaged in $b$.

It is important to notice that there are two mechanisms to increase $\sigma_{tot}(s)$ and $\sigma_{el}(s)$: by expanding the cross-section disk, i.e., by increasing $R(s)$, or by increasing matter density, with an approach to the black disk, by increasing $\overline{\Omega}(s)$. Moreover, the ratio $\sigma_{el}(s)/\sigma_{tot}$ is just a function of $\overline{\Omega}(s)$,

\begin{equation}
\frac{\sigma_{el}(s)}{\sigma_{tot}(s)} = \frac{1}{2} \left( 1 - e^{- \overline{\Omega} (s)}\right),
\label{eq:sigmael_tot}
\end{equation}

\begin{figure}[htbp]
  \begin{center}
   \includegraphics[width=0.5\textwidth]{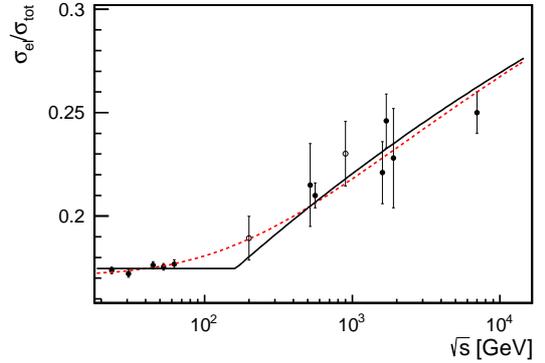}
   \caption{Elastic over total cross-section as a function of $\sqrt{s}$. The full line shows the best fit within this model. The data points used to perform the fit are shown as filled circles (see text for details). Moreover, notice the transition at $\sqrt{s} \sim 160$ GeV. The red (dashed) line shows the same model with smooth transitions.}
   \label{fig:2}
 \end{center}
\end{figure}

Let us consider all the data from ISR to LHC ($\sqrt{s} \sim 20 \text{ GeV to } 7 \text{ TeV}$) \cite{eltotdata1,eltotdata2,eltotdata3,eltotdata4,TOTEM}, see figure \ref{fig:2}. The data points at $\sqrt{s} = 200$ GeV and $900$ GeV, which appears as unfilled circles in the plots, were left out of the present analysis as the cross-sections were obtained through the extrapolation of the low energy data \cite{eltotdata3}. However their impact to the fits was checked and found to be negligible. There are clearly two different regions. Up to $\sqrt{s} \sim 100$ GeV $\sigma_{el}/ \sigma_{tot}$ is constant while above this ratio grows as $\ln \sqrt{s}$. 
Relating the number of collisions to $\overline{\Omega}$ making the assumption

\begin{equation}
\overline{\nu}(s) = e^{k \overline{\Omega}(s)} \simeq 1 + k \overline{\Omega}(s) + ...
\label{eq:nstrings}
\end{equation}

($\overline{\Omega} \rightarrow 0$, $\overline{\nu} \rightarrow 1$ and $\overline{\Omega} \rightarrow \infty$, $\overline{\nu} \rightarrow \infty$ ), and using (\ref{eq:two}) we obtain

\begin{equation}
\overline{\Omega}(s) = \frac{2}{k} \lambda Y,
\label{eq:density}
\end{equation}

$\overline{\Omega}$ does increase with $\ln \sqrt{s}$ and a good fit to the region $\sqrt{s} > 100$ GeV is thus obtained. ($k = 5.52 \pm 0.15$). Note that the relation $\overline{\nu} \sim \overline{\Omega}$, instead of (\ref{eq:nstrings}), does not agree with data for the observed value of $\lambda$.

If we look now at the ISR data ($20-60$ GeV) - left hand side of figure \ref{fig:2} - we see that (\ref{eq:nstrings}) and (\ref{eq:density}) obviously do not apply, as 

\begin{equation}
\overline{\Omega} = \text{const} = 0.4298 \pm 0.0024
\label{eq:Omegaconst}
\end{equation}

Thus there are two regimes: one, corresponding to approach to the black disk, where equation (\ref{eq:density}) applies; the other, corresponding to a constant value of $\sigma_{el} / \sigma_{tot}$, Geometric Scaling (GS) \cite{ref8_1,ref8_2}. Note that asymptotically as $\overline{\Omega}(s) \rightarrow \infty$ one recovers GS and the results of \cite{ref9_1,ref9_2}.

\begin{figure}[htbp]
  \begin{center}
   \includegraphics[width=0.5\textwidth]{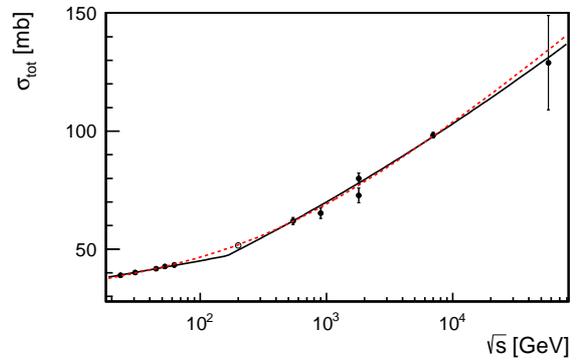}
   \caption{Total cross-section as a function of $\sqrt{s}$. The full line is the model best fit. The data points used to perform the fit are shown as filled circles (see text for details). The red (dashed) line shows the same model with smooth transitions.}
   \label{fig:3}
 \end{center}
\end{figure}

The radial expansion is controlled by the function $R(s)$. We shall consider two terms: $R_0$, which is a constant presumably related to the valence quark content of beam and target particles, and a $\beta \log \left( s/s_0 \right)$ term, inspired on the universal behaviour first predicted by Heisenberg \cite{Heisenberg1,Heisenberg2}, and compatible with the Froissart bound \cite{Froissart,Martin}, such that

\begin{equation}
R(s) = R_0 + \beta \log \left( \frac{s}{s_0} \right).
\label{eq:radius_term}
\end{equation}

In figure \ref{fig:3} we show our fits to data on $\sigma_{tot}(s)$ \cite{eltotdata1,eltotdata2,eltotdata3,eltotdata4,XsecPAO}, with the expectations at LHC, for pp $\sigma_{tot}(\sqrt{s} = 14\ \text{TeV})$. The preferred parameters are $R_0 = 2.52 \pm 0.02$ GeV$^{-1}$, for pp collisions, and $\beta=0.067\pm0.002$ GeV$^{-1}$. In this fit $\sqrt{s_0} = 10$ GeV, and the $k$-parameter was fixed to the value obtained in figure \ref{fig:2}.

\begin{figure}[htbp]
  \begin{center}
   \includegraphics[width=0.5\textwidth]{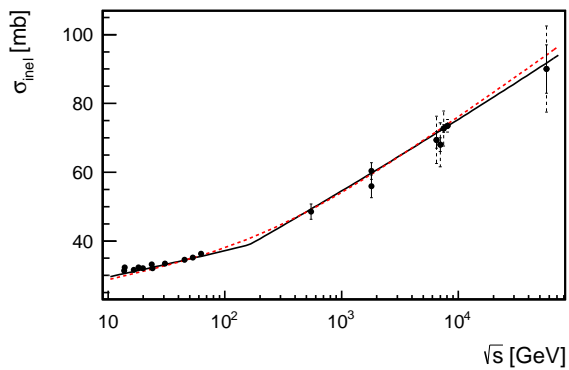}
   \caption{Inelastic cross-section as a function of $\sqrt{s}$. The full line is obtained from equation \ref{eq:greydiskinel} with all parameters fixed from the previous fits. The new experimental points at $\sqrt{s} = 7$ TeV from CMS \cite{XsecCMS}, ATLAS \cite{XsecATLAS}, ALICE \cite{XsecALICE} and TOTEM \cite{TOTEM} (left to righ in the plot) are shown together with the statistical (full line) and systematic errors (dotted). The point at $\sqrt{s} = 57$ TeV is from the Pierre Auger Collaboration \cite{XsecPAO}. The red (dashed) line shows the same model with smooth transitions.}
   \label{fig:4}
 \end{center}
\end{figure}

In figure \ref{fig:4} we present our expectations on $\sigma_{inel}(s)$ and compare it to the available data \cite{ineldata1,eltotdata4}, including the new LHC measurements at $\sqrt{s} = 7$ TeV \cite{XsecATLAS,XsecCMS,XsecALICE,TOTEM} and new data from the Pierre Auger Collaboration at $\sqrt{s} = 57$ TeV ($E \sim 10^ {18}$ eV) \cite{XsecPAO}. All the inelastic cross-section data is well reproduced without any new free parameters, which indicates that both the model and the data are correlated with the previous fitted cross-sections by unitarity. It is interesting to note also that with this model we obtain essentially the same cross-section values that in more sophisticated models \cite{Block1,Block2}.


The Pierre Auger Collaboration has recently reported results at energies $\sim 10^{19}$ eV\footnote{$\sqrt{s} \sim 100\ {\rm TeV}$} which are not easily explained in the framework of any \emph{standard model} or smooth evolution \cite{augerxmax,augermuons}. The average depth of the shower maximum, $X_{max}$ and its fluctuations ($RMS$) have a sudden change (in the case of the $RMS$ corresponds to a dramatic decrease from a value of $60\ {\rm g\,cm^2} $ to almost $20\ {\rm g\,cm^2} $ in less than an energy decade), while the number of muons at ground, although much higher than expected, has a smooth evolution with energy, within the present statistics. These results are not compatible, as shown by several authors (see for example \cite{Wilk}), with the most straightforward explanation: the composition of the primary cosmic rays changes suddenly from proton to iron.
An alternative explanation for a composition change is a fast increase of proton-Air cross-section above a certain energy threshold\footnote{There are new analyses exploring the shape of the longitudinal profile \cite{USPV} that hopefully will be able to separate in a near future the cross-section from the hadronic models effects.}. This approach has been discussed already by several authors \cite{augerUlrich,UlrichPhD,Wilk} in an empirical way. Here we purpose that a sudden change in the cross-section can be explained within the framework of a grey disk model. Indeed, in our model such an increase could only be obtained  by a sharp increase in $\overline{\Omega}(s)$, and $\sigma_{el}/\sigma_{tot}$, which remain even at LHC, far from black-disk limit (for a recent discussion on $\sigma_{el} / \sigma_{tot}$ see \cite{Fagundes}). 

\begin{figure}[htbp]
  \begin{center}
   \includegraphics[width=0.5\textwidth]{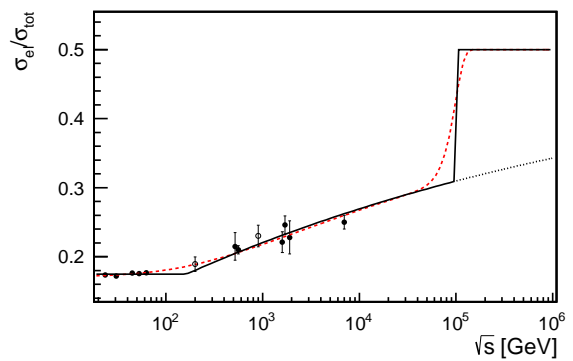}
   \caption{$\sigma_{el} / \sigma_{tot}$ as a function of $\sqrt{s}$, plotted up to cosmic ray energies. The full (black) line shows the model with non-regular functions while the red (dashed) line represents the same model with a smooth transition (see text for details).}
   \label{fig:5}
 \end{center}
\end{figure}
\begin{figure}[htbp]
  \begin{center}
   \includegraphics[width=0.5\textwidth]{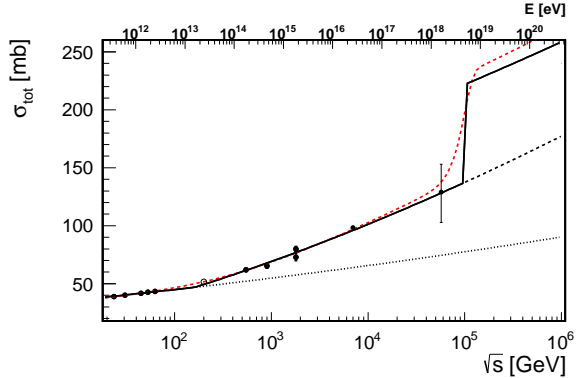}
   \caption{Total cross-section as a function of $\sqrt{s}$ up to cosmic ray energies. An abrupt increase is possible without violation the Froissart limit and can accommodate the Pierre Auger Observatory data at energies above $100$ TeV \cite{augerxmax}. The full (black) line shows the model with non-regular functions while the red (dashed) line represents the same model with a smooth transition (see text for details). The upper scale is the corresponding energy of the cosmic ray.}
   \label{fig:6}
 \end{center}
\end{figure}

If we assume that at $\sqrt{s} \approx 100$ TeV $\Omega(s) \rightarrow \infty$ (or $\sigma_{el}/\sigma_{tot} \rightarrow 1/2$) we have a sharp transition at $\sqrt{s} \approx 100$ TeV to a black-disk, see figures \ref{fig:5} and \ref{fig:6}. This increase could be originated due to an increase of $\lambda$ which implies an increase of multiplicity as well as rise in cross-section.

Note that the Froissart bound is not affected. Our curve, in the left hand of figure \ref{fig:5}, is below a factor of the order of $100$ in comparison with Froissart bound limit \cite{Froissart}. Hence, the fast evolution to a black disk at sqrt(s) ~ 100 TeV would allow to have a very fast increase in the proton-proton cross-section to a value up to $80\%$ more. In any case, the sharp increase of the cross-section should be seen as a local asymptotical behaviour.

Let us now look at our model in a representation making use exclusively of regular functions, without a discontinuity at $\sqrt{s} \approx 160$ GeV and another at $\sqrt{s} \approx 100$ TeV (see Figs. \ref{fig:2} and \ref{fig:4}). We now consider $k$ (see equation \ref{eq:density}) a continuous function of $\sqrt{s}$, such that $k = A Y$ at low energy and $k = \text{const} = B$ above $\sqrt{s} \approx 160$ GeV and behaving as a Fermi function around $\sqrt{s} \approx 100$ TeV:

\begin{equation}
k = \underbrace{\left( \frac{AY}{e^\frac{Y-Y_1}{\Delta}+1} + \frac{B}{e^{-\frac{Y-Y_1}{\Delta}}+1} \right)}_\text{first transition} \underbrace{\cdot \frac{1}{e^\frac{Y-Y_2}{\delta}+1}}_\text{second transition}
\label{eq:smoothk}
\end{equation}

where A,B, $\Delta$ and $\delta$ are free parameters. Using equation \ref{eq:smoothk} as an input for equation \ref{eq:density} it is possible to obtain a representation of $\sigma_{el}/\sigma_{tot}$ (eq. \ref{eq:sigmael_tot}). The red (dashed) curves in Figs. \ref{fig:2} to \ref{fig:6}, correspond to our continuous representation ($A = 0.89 \pm 0.04$; $B = 5.47 \pm 0.21$; $\Delta = 1.36 \pm 0.22$; and $\delta$ was fixed to $0.2$). The change occurs mostly in the region $\sqrt{s} \approx 160$ GeV. It would be nice to have more information in that particular region.

The formation of a black disk in the way to asymptotia, as advocated here, has been discussed in the literature for quite some time (see, for instance \cite{x} and \cite{y}). However there are alternative approaches (see \cite{z}), where the asymptotia occurs via grey disk formation.

\section*{Final Remarks}

We have shown that a simple grey disk model, characterized only by an effective radius and an opacity, can be used to described consistently all the available data on proton-proton cross-sections (total, elastic and inelastic) from $\sqrt{s} \sim 10\ {\rm GeV} - 57\ {\rm TeV}$.

The drastic change in the ratio of elastic to total cross-section in proton-proton interactions around $\sqrt{s} \sim 100$ GeV is a well established experimental fact that has not so far a reasonable explanation. This fact has been disregarded by many authors, perhaps because it is only noticeable in this ratio (and not in the cross-section separately) and in a proper scale. In this model, we propose that such a transition could be explained by an increase of the density $\overline{\Omega}(s)$. It would be interesting to have experimental information in the $\sqrt{s} \simeq 100$ GeV region, where the transition from one regime to another is supposed to occur.

At much higher energies ($\sqrt{s} \simeq 100$ TeV) is not yet clear if a new drastic change in the cross-section occurs but, as discussed above, such scenario should be taken seriously. We have shown that such a transition can be accommodated without violating the Froissart bound. Notice that with a normal growth of the cross-sections, $\sigma_{tot} \sim \log^2 (s/s_0)$, one should reach the black disk limit only at very high energy (for instance, $\sigma_{el}/\sigma_{tot} \simeq 0.49$ occurs for $\sqrt{s} \simeq 10^{19}$ GeV).
In general a rapid rise in cross-sections is associated to opening of new channels. However, the opening of new flavour channels $-$ contrary to what is observed in $e^+e^-$ cross-section $-$ was never detected in hadron cross-sections. But in the present case we have perhaps not a new flavour but a new world of physics (for instance a new scale for compositeness).


\section*{Acknowledgments}
We would like to thank C. Pajares, L. Cazon, S. Andringa, F. Navarra and J. Alvarez-Mu\~{n}iz for suggestions and encouragement. R.C. wants to acknowledge the financial support given by FCT, Funda\c{c}\~{a}o para a Ci\^{e}ncia e Tecnologia (SFRH/BPD/73270/2010).


\bibliographystyle{elsarticle-num}
\bibliography{Bib-xsecGreyModel}

\end{document}